\documentclass[12pt,preprint]{aastex}

\slugcomment{Submitted to ApJL}

\shorttitle{Expansion velocities and core masses of bright PNs in the
Virgo cluster} \shortauthors{Arnaboldi, M. et al.}

\begin{document}


\title{Expansion Velocities and Core Masses of Bright Planetary
Nebulae in the Virgo Cluster}


\author{Magda Arnaboldi\altaffilmark{1,2}, Michelle
Doherty\altaffilmark{1}, Ortwin Gerhard\altaffilmark{3}, Robin
Ciardullo\altaffilmark{4}}
\author{J. Alfonso L. Aguerri\altaffilmark{5}, John
J. Feldmeier\altaffilmark{6}, Kenneth C. Freeman\altaffilmark{7},
George H. Jacoby\altaffilmark{8}}


\altaffiltext{1}{European Southern Observatory, Garching, Germany;
marnabol@eso.org, mdoherty@eso.org} 
\altaffiltext{2}{INAF, Observatory of Turin, Turin, Italy}  
\altaffiltext{3}{Max-Planck-Institut f\"ur extraterrestrische Physik, Garching,
Germany; gerhard@mpe.mpg.de }
\altaffiltext{4}{Dept. of Astronomy and Astrophysics, Pennsylvania
State University, University park, PA; rbc@astro.psu.edu}
\altaffiltext{5}{Instituto de Astrofisica de Canarias, Tenerife,
Spain; jalfonso@ll.iac.es}
\altaffiltext{6}{Dept. of Physics and Astronomy, Youngstown
State University, Youngstown, OH; jjfeldmeier@ysu.edu}
\altaffiltext{7}{Mount Stromlo Observatory, Research School of
Astronomy and Astrophysics, ACT, Australia;kcf@mso.anu.edu.au}
\altaffiltext{8}{WYIN Observatory, Tucson, AZ; jacoby@noao.edu}


\begin{abstract}
  The line-of-sight velocities and [OIII] 5007 \AA\ expansion
  velocities are measured for 11 planetary nebulae (PNs) in the Virgo
  cluster core, at 15 Mpc distance, with the FLAMES spectrograph on
  the ESO VLT.  These PNs are located about halfway between the two
  giant ellipticals M87 and M86. From the [OIII] 5007 \AA\ line
  profile widths, the average half-width at half maximum expansion
  velocity for this sample of 11 PNs is $ \bar{v}_{HWHM} = 16.5$
  kms$^{-1}$ (RMS = 2.6 kms$^{-1}$). We use the PN subsample bound to
  M87 to remove the distance uncertainties, and the resulting [OIII]
  5007 \AA\ luminosities to derive the central star masses. We find
  these masses to be at least $0.6$ M$_\odot$ and obtain PN observable
  life times $t_{PN} < 2000 $ yrs, which imply that the bright PNs
  detected in the Virgo cluster core are compact, high density
  nebulae. We finally discuss several scenarios for explaining the
  high central star masses in these bright M87 halo PNs.
\end{abstract}


\section{Introduction}

Since the discovery of free-floating intracluster planetary nebulae
(ICPNs) in the Virgo cluster \citep{Arnaboldi+96}, extensive imaging
and spectroscopic observations were carried out to determine their
projected phase space distribution and the fraction of diffuse cluster
light not bound to Virgo galaxies. To enlarge the sample of more than
40 ICPN line-of-sight (LOS) velocities available from
\citet{Arnaboldi+03,Arnaboldi+04}, we have obtained new PN spectra
with FLAMES at the ESO VLT \citep{Doherty+08}. The high spectral
resolution of the new data allows us for the first time to measure
expansion velocities for the [OIII] nebula shells of 11 PNs in the
Virgo cluster core, nearly half way between M87 and M86.  The
expansion velocity of a planetary nebula (PN) is one of the most
important parameters determining its evolution, but currently it is
known only for a few hundred Galactic PNs, mostly bright objects in
the Milky Way Disk \citep{gesickizijlstra00}, and for a few tens in
the Magellanic Clouds and M31. When interpreted using dynamically
evolving nebular models \citep[e.g.][]{Schoenberner+05}, PN shell
expansion velocities provide reliable extimates of PN dynamical ages
and distance estimates for Galactic PNs.

In this letter, we give a brief summary of our Observations in
Section~\ref{obser}. In Section~\ref{PNexp}, we present the PN [OIII]
5007 \AA\ half-width half maximum velocity measurements $v_{HWHM}$,
and $m(5007)$ magnitudes as defined by \citet{Jacoby89}. In
Section~\ref{M87PNLF} we build the PN luminosity function (PNLF) for
the spectroscopically confirmed PN sample in Virgo and for the
subsample bound to the M87 halo.  In Section~\ref{outerradius}, we
then estimate the outer radii of the M87 halo PNs from the observed PN
expansion velocities and their visibility lifetimes.  We finally
discuss in Section~\ref{fdiscuss} the remaining distance
uncertainties, the shape of the PNLF in the M87 halo, and the
mechanisms that may be responsible for the large core masses of the
brightest PNs in the halo of M87.

\section{Observations\label{obser}}

Data were acquired in service mode (22 hrs, 076.B-0086 PI: M.
Arnaboldi), with the FLAMES spectrograph at UT2 on VLT in the
GIRAFFE+MEDUSA configuration. We used the high-resolution grism HR
504.8, covering a wavelength range of 250 \AA\, centered on 5036 \AA\,
and a spectral resolution of $\sim 20000$. The redshifted [OIII]
emissions of PNs in the Virgo cluster core fall near the center of the
grism transmission. With this setup, the instrumental broadening of
the arc lines is FWHM = 0.29 \AA\ or 17 kms$^{-1}$, and the error on
the wavelength measurements is 0.0025 \AA\ or 150 ms$^{-1}$
\citep{Royer+02}. A total of three FLAMES plate configurations were
produced for the Virgo core fields F4 and F7 surveyed by
\citet{Feldmeier+03}. The exposure times were based on the
signal-to-noise ratio (S/N) estimate for detecting the [OIII] 5007
\AA\ line flux of $4.2 \times 10^{-17} $ergs~cm$^{-2}$~s$^{-1}$, i.e.,
$m(5007) = 27.2$, with S/N $\sim$ 5. The data reduction was carried
out with the GIRAFFE pipeline, for the CCD prereduction, fiber
identification, wavelength calibration, geometric distortion
corrections, co-addition, and extraction of the final one-dimensional
spectra. Data analysis and the velocity measurements are presented in
Doherty et al. (2008). A total of 12 PNs were confirmed
spectroscopically.  In the extracted PN spectra, the [OIII] 5007 \AA\
emissions of 11 PNs have measured FWHM between $0.4$ and $0.7$ \AA ;
the comparison with the arclines' FWHM ($0.29$ \AA) indicates that
these [OIII] emission lines are resolved. Monte Carlo simulations give
an error on the line width measurements of $0.0025$\AA\ for a S/N per
pixel of $\sim 15$ \citep{Royer+02}.

\section{PN expansion velocities\label{PNexp}}

The spectroscopic expansion velocity measured from the [OIII] 5007\AA\
line profile appears to be representative for the material velocities
associated with the PN bright rim\footnote{The bright rim is the thin
  shell well behind the outer shock, enclosing the wind-blown cavity
  of a PN.}, and is systematically lower than the
expansion speed of the PN shell's outer radius, according to 1D
hydrodynamical simulations \citep{Schoenberner+05}. The shell
expansion velocities are associated with fainter structures at larger
radii, which have been measured only for Galactic PNs
\citep{Corradi+07}.

\begin{figure}
\epsscale{1.0} 
\plotone{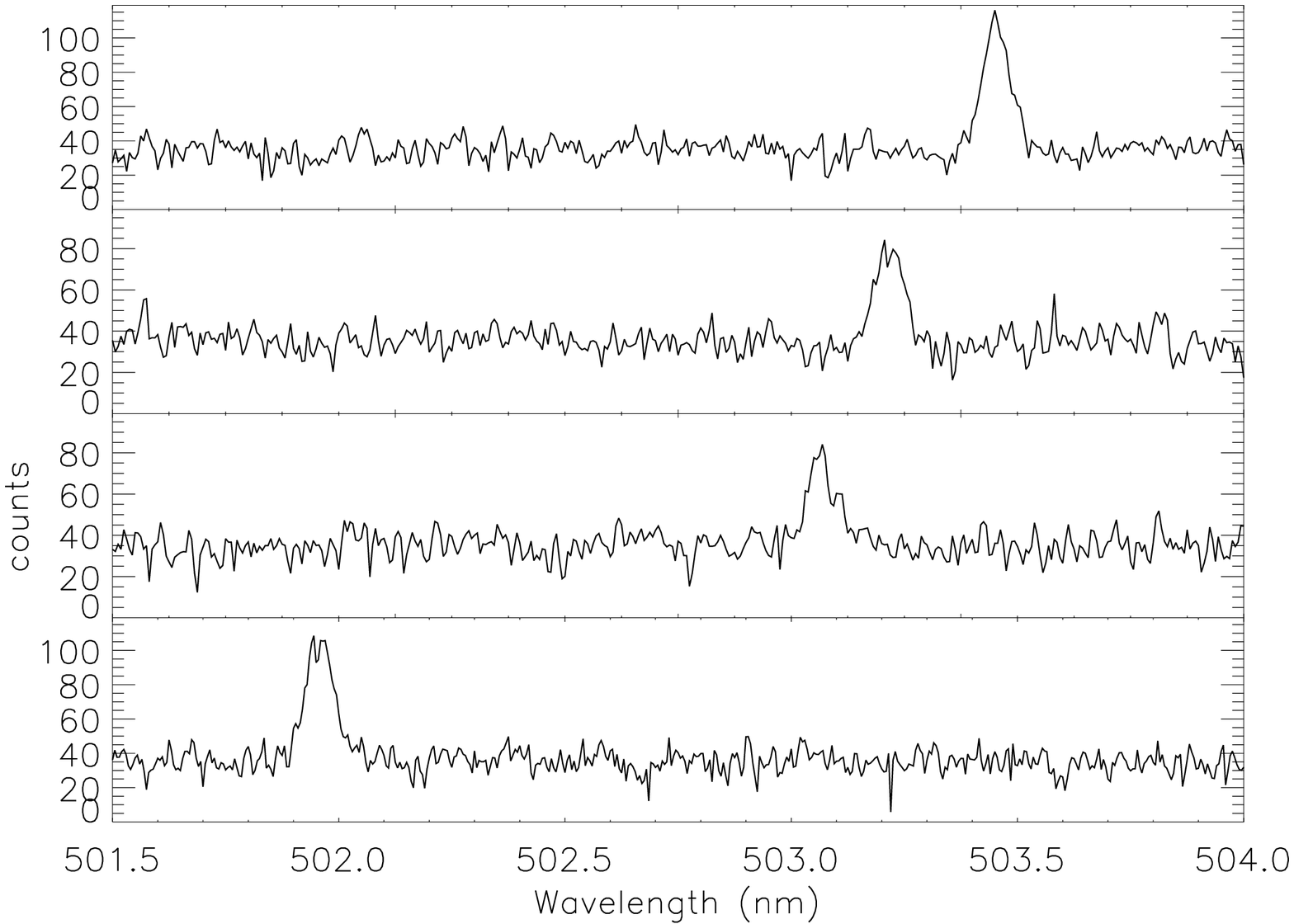}
\caption{Resolved [OIII] 5007 \AA\ emission lines of 4
spectroscopically confirmed PNs. \label{fig0}}
\end{figure}

From the FLAMES PN spectra, we are able to measure the spectroscopic
expansion velocities from the resolved [OIII] 5007\AA\ emissions for
11 PNs in the \citet{Doherty+08} sample; high S/N spectra for 4 PNs
are shown in Fig.~\ref{fig0}. We use the ``half-width at half
maximum'' of the [OIII] 5007 \AA\ line, corrected for the instrumental
half width, $v_{HWHM}$, as measurement of the spectroscopic expansion
velocity for the PN bright rim. According to \citet{Schoenberner+05}
and \citet{Corradi+07}, these $v_{HWHM}$ measurements must be
multiplied by a factor of about two to get an estimate of the PN's
true expansion velocity.  When it is possible, we also measure the
spectroscopic expansion velocities for [OIII] 4959 \AA. When double
peaks or secondary peaks are present in the resolved line profiles,
their counts are consistent with the noise.  The average [OIII]
spectroscopic expansion velocity for this PN sample is $\bar{v}_{HWHM}
= 16.5$ kms$^{-1}$, with an RMS dispersion of 2.6 kms$^{-1}$. Both
values are significantly smaller than those determined for samples of
PNs observed in the Galactic bulge \citep{gesickizijlstra00}.  The
Virgo PNs might represent the [OIII] brightest part of the Galactic
bulge population, or they might come from an entirely different
population; this issue requires further work.

\begin{figure}
\epsscale{1.1} \plotone{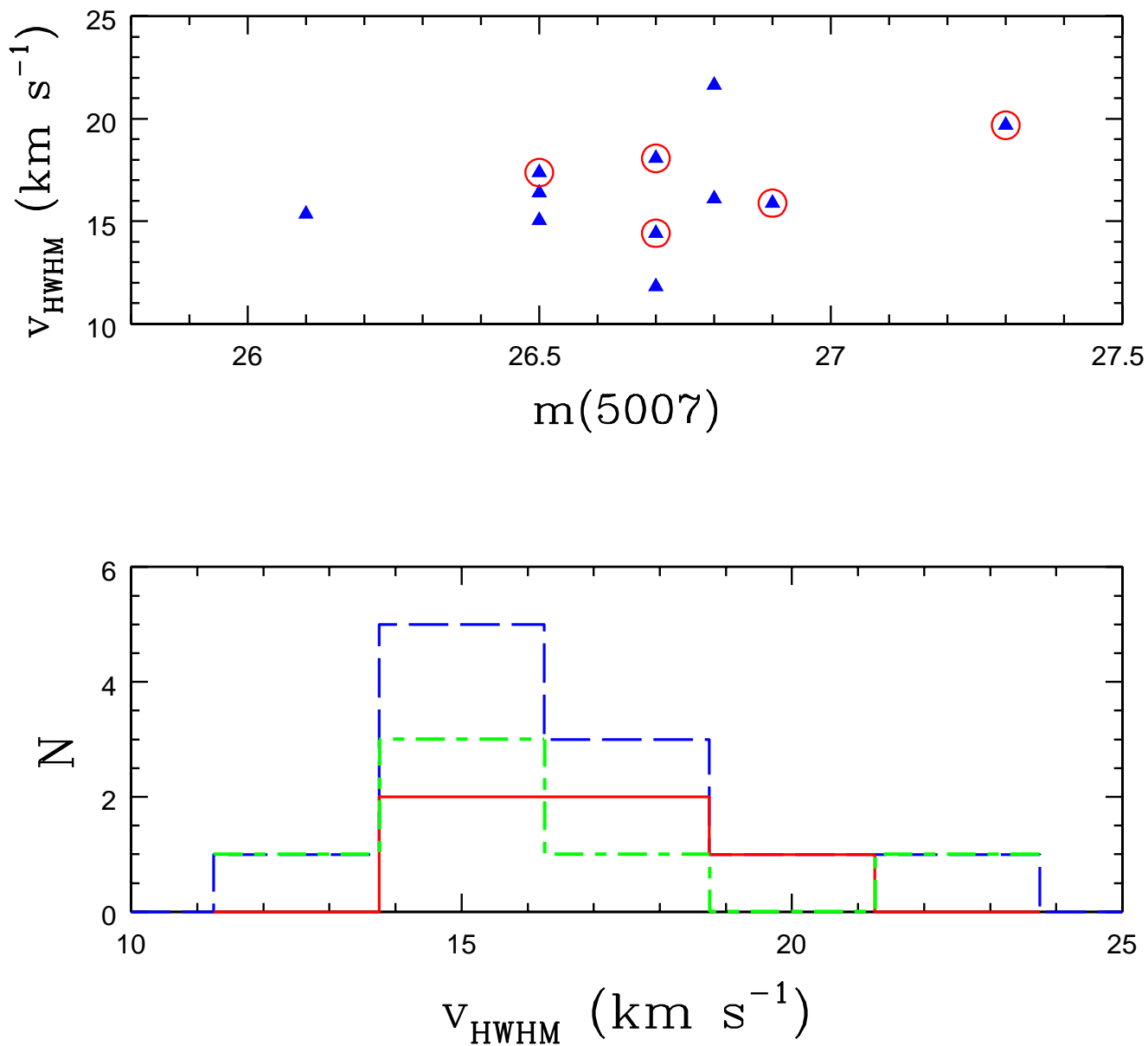}
\caption{Top panel - [OIII] spectroscopic expansion velocities
$v_{HWHM}$ vs. $m(5007)$ magnitudes for the PNs from the \citet{Doherty+08}
sample; the circled symbols indicate the M87 PNs. Bottom panel -
histograms of $v_{HWHM}$ measured for all PNs in the \citet{Doherty+08}
sample (dashed blue lines), for the subsample bound to the M87 halo 
(continuous red lines), and for the PNs not bound to M87 (dash-dotted 
green lines). The average $v_{HWHM}$ for the entire sample is 16.5
kms$^{-1}$. \label{fig1}}
\end{figure}

Based on LOS velocities, we divide the current dataset in two
subsamples: 5 PNs are associated with the halo of M87, i.e.  belong to
a narrow velocity peak in the LOS velocity distribution (LOSVD)
centered at the M87 systemic velocity, with $RMS = 78$ kms$^{-1}$
\citep{Doherty+08}, and 6 are ``free-flying'' cluster PNs, with $300 -
1500$ kms$^{-1}$ velocity differences. The distributions of the
$v_{HWHM}$ measurements for the two subsamples are similar within the
limited statistics, see Fig.~\ref{fig1}; the KS-test gives only a 55\%
probability that the two subsamples are different. The $v_{HWHM}$ vs.
m(5007) plot in the top panel of Fig.~\ref{fig1} shows that the PNs
within 0.7 mags of the bright cut-off have $v_{HWHM} < 20$ kms$^{-1}$.
This is in agreement with the predictions of the spatially integrated
line profile $v_{HWWM}$ vs. $m(5007)$ computed with the nebular 1D
hydrodynamics code of \citet[][and in prep.]{Schoenberner+07} for
central star masses in the range $ 0.696 - 0.625$~M$_\odot$.

When a PN's distance is known, its expansion velocity and size can be
used to determine its dynamical age. This age, in turn, can be
combined with information about the central star's effective
temperature to yield an estimate of the core mass
\citep{gesickizijlstra07}.  In our case we cannot measure the outer
radii for the Virgo PNs, as they are unresolved at a distance of 15
Mpc.  Our approach is then the reverse: we shall estimate the central
star masses from the [OIII] 5007 \AA\ fluxes, derive the PN lifetimes
$t_{PN}$ from central star evolutionary tracks and nebular 1-D
hydrodynamical models, and infer their physical dimensions as $r_{PN}
= v_{exp}\times t_{PN}$, where $v_{exp}= 2 \times v_{HWHM}$
(Schoenberner et al., in prep.).

\section{The M87 halo PNLF\label{M87PNLF}}

\begin{figure}
\epsscale{1.1} \plotone{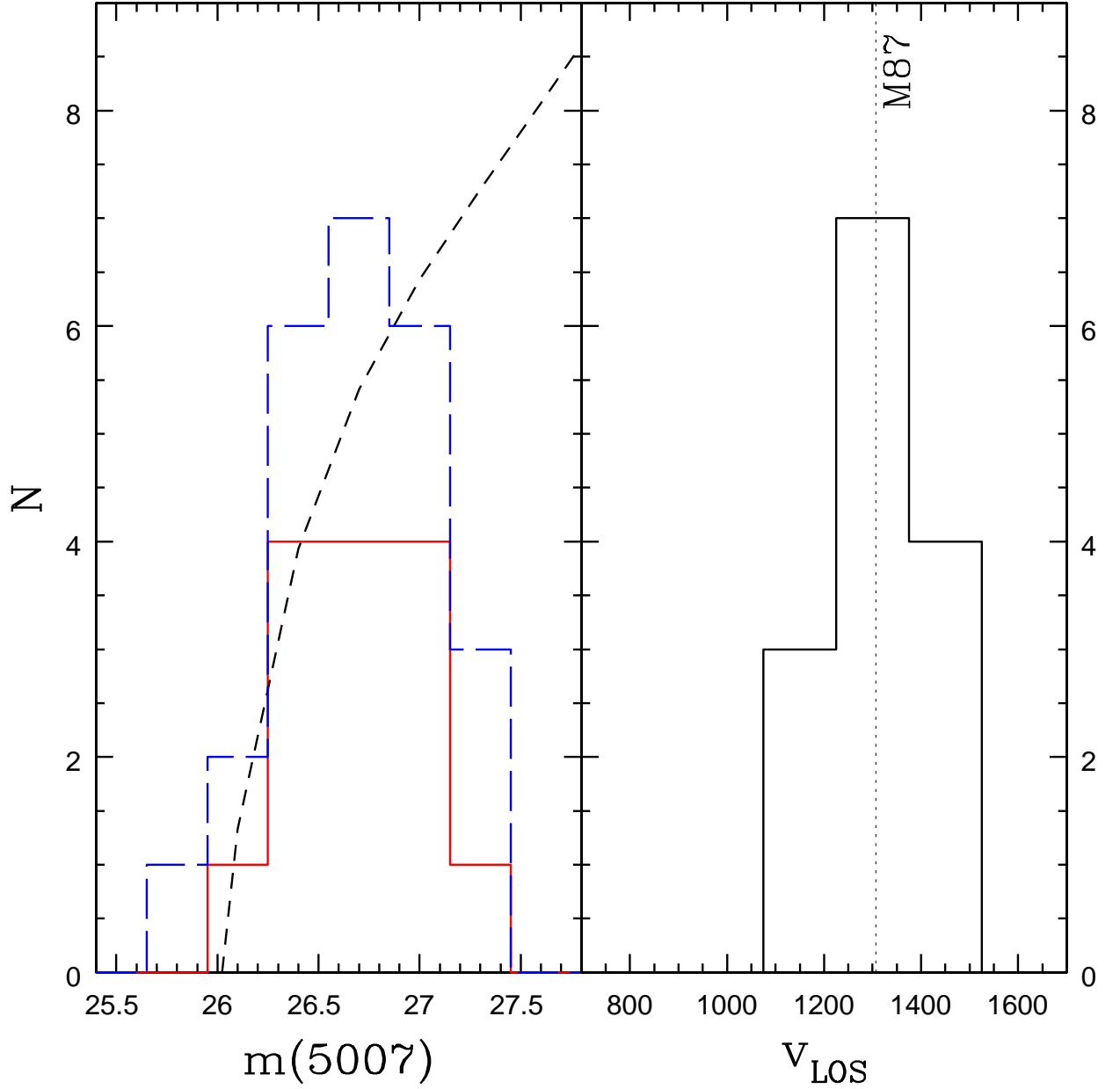}
\caption{Left panel: PNLF in the surveyed fields in 0.3 mag bins. Long
dashed blue line: PNLF for the entire spectroscopically confirmed
sample of PNs in the Virgo cluster core at the position of the F3, F7
and F4 fields of \citet{Feldmeier+03}. Red continuous line: PNLF for
the PNs bound to the M87 halo. Short dashed black line:
\citet{Ciardullo+89} analytical formula for the PNLF with $m^* =
26.0$. Right panel : LOS velocity distribution for the 14 PNs bound to
the M87 halo. The systemic velocity of M87 is also
indicated. \label{fig3}}
\end{figure}

The PNLF for the spectroscopically confirmed PNs in the Virgo core
region around M87 is shown in Figure~\ref{fig3}. The brightest PNs
have $m(5007)$ in the range $25.7 - 26.5$.  PNs with similar
magnitudes were also detected by \citet{Ciardullo+98} in the outer
regions of M87. Based on a sample of 338 PNs, \citet{Ciardullo+98}
determined a brightening of the PNLF cutoff of 0.37 mag for the PN
subsample at $R > 4'$ with respect to the sample inside $R \leq 4'$.
They interpreted the brightening of the PNLF in the M87 outer halo as
due to a population of Virgo ICPNs filling the elongated volume of the
Virgo cluster (up to 4 Mpc along the LOS).  The number of foreground
PNs would be roughly proportional to the area of the field, and
therefore be largest in the outer regions of the surveyed field.

In our spectroscopic PN sample we can bypass the distance ambiguity by
selecting the subsample of 14 PNs\footnote{5 PNs from the
  \citet{Doherty+08} sample and 9 from the F3 field from
  \citet{Arnaboldi+04}} bound to M87. These PNs are associated with a
narrow peak in the LOSVD centred at the systemic velocity of M87, see
Fig.~\ref{fig3}. Their average velocity $v_{LOS}\!=\!1306$ kms$^{-1}$
and the $RMS\!=\!  117$ kms$^{-1}$.  The empirical PNLF for the M87
halo PNs is shown in Fig.~\ref{fig3}: their $m(5007)$ magnitudes are
in the range $26.2 - 27.2$, indicating a slightly brighter cut-off
than for the PN population in the central $R \leq 4'$ region of M87.
However we can now say that this is an intrinsic property of the PN
halo population, as we now know that these 14 PNs are all at the
distance of M87 within $\sim 100$ kpc (see also \S6).

\section{PN core masses, visibility time scales and 
outer radius\label{outerradius}}

From the measured $m(5007)$ of the M87 halo PNs, we estimate the
central star total luminosity. Both models and observations indicate
that no more than 10\% of the central star's total luminosity comes
out in this line \citep{Jacoby89}. Therefore the intrinsic luminosity
of the post-AGB star that powers an $m(5007) = 26.2$ PN at 14.5 Mpc
must be $> 6930$ L$_\odot$ and, based on the evolutionary tracks of
\citet{Bloecker95}, we thus obtain a central core mass larger than $
0.6~$M$_\odot$. Using a 1-D hydrodynamics code and the stellar
evolution tracks of \citep{Bloecker95}, \citet{Schoenberner+07}
computed the nebular physical parameters and evolution of $m(5007)$ as
function of nebular age, from near the AGB phase to the white dwarf
cooling tracks. The nebular tracks of
\citet[][Fig.~15]{Schoenberner+07} that reach the brighest 0.5 mag of
the PNLF, and have a central core mass of $> 0.60$M$_\odot$ as the
Virgo PNs, have very short visibility lifetimes, $t_{PN} < 2.0 \times
10^3$ yrs. From our measurements and $v_{exp}= 2 \times v_{HWHM}$, we
can then determine the outer radii $r_{PN}$ of the brightest PNs in
the M87 halo to be $ \sim 0.07$ pc. The nebular shells of these PNs
are compact and similar to those observed for the brightest ($\log L
\ge 3.8$) Galactic Bulge PNs.

\section{Discussion\label{fdiscuss}}

{\it Distance uncertainties and PNLF brightening -}
We briefly consider the question whether the bright PNe in the
velocity range bound to M87 could be foreground objects. 
The mean heliocentric velocity of the Virgo cluster is $<v_\odot>_{VC}
= 1050 \pm 35$ kms$^{-1}$ \citep{Binggeli+93}. M87 is redshifted
by about 300 kms$^{-1}$ ($v_{M87} = 1307$ kms$^{-1}$) with respect to
the Virgo cluster mean velocity,  and is falling into Virgo from
in front \citep{Binggeli+93} towards M86 \citep{Doherty+08}.

PNs at the cutoff $m^\ast=26.33$ with apparent magnitudes 26.2-26.0
would be between $\sim 1-3$ Mpc in front of M87.  Diffuse light stars
at these locations would be in the infall region towards Virgo where
the infall velocities are of order 1000 kms$^{-1}$ and vary rapidly with
distance \citep[][and references therein]{Mohayee+05}.  On the other
hand, radial velocities of $\sim 1300$ kms$^{-1}$ would be expected for
foreground objects $\sim 6$ Mpc in front of the Virgo core.  It is
thus very unlikely that a distribution of foreground PNe should be
observed in our fields at exactly the systemic velocity of M87, with a
dispersion of only $\sim 100$ kms$^{-1}$, and less likely still that this
population should be projected onto the M87 halo inside 150 kpc but
not be observed in the adjacent Virgo core region covered by our data.

While the Virgo cluster has a significant depth ($\sim\pm2$Mpc), it is
unlikely that the diffuse light and ICPN distribution are similarly
elongated.  The search for ICPNs outside the Virgo cluster core has
given negative results (Castro et al. 2008, in prep.), and the deep
photometry of \citet{Mihos+05} shows that the diffuse light is mostly
associated with the giant elliptical galaxies, M87, M86 and M84,
whereas its surface brightness decreases sharply at larger radii.
Observations of the diffuse light in $z=0.25$ galaxy clusters also
show that it is more centrally concentrated than the cluster galaxies
\citep{Zibetti+05}, in agreement with cosmological simulation of
cluster formation \citep{Murante+07}. Thus we can safely conclude that
the selected M87 halo PNs are bound to M87, and that their bright
[OIII] magnitudes are an intrinsinc property of this PN population.

{\it Mechanism leading to large core masses in the M87 halo PNs - }
The initial mass-final mass relation (IFMR) for solar metallicity
stars predicts that the PN progenitors with $\sim 2.2$ M$_\odot$ give
final core masses of $0.62$ M$_\odot$ \citep[see][]{Ciardullo+05,
  Buzzoni06}.  Turnoff masses of $\sim 2$~M$_\odot$ belong to
populations with ages $\sim 1$ Gyr \citep{Iben89}. The question is
whether such populations exist in the M87 halo or in the Virgo diffuse
stellar light. Observations in the V, I bands of the stellar
population in a 11.39 arcmin$^2$ field halfway between M87 and M86
were carried out with the Advanced Camera for Survey (ACS) and the
HST. This location falls within the F4 Virgo field of
\citet{Feldmeier+03} and is included in the area surveyed by
\citet{Doherty+08}.  At this location, \citet{Williams+07} detected
some $\sim 5300$ intracluster red giant branch stars (IRGB); from the
color magnitude diagram (CMD), they estimated the age and metallicity
distribution of the parent stellar population. In this region, 70\% -
80\% of the Virgo IRGB are old ($> 10$ Gyr), and span a wide range of
metallicities ($ -2.3 < [M/H] < 0.0$), with a mean value of $[M/H]
\sim -1.0$.

From the number of PNs within 0.5 mag of the PNLF cut-off, $N_{0.5}$,
we can determine how much luminosity would be present in an
intermediate age population and compare it with the measured surface
photometry of \citet{Mihos+05} in the surveyed region and the
\citet{Williams+07} fit to the IRGB CMD. If we assume the analytical
formula of \citet{Ciardullo+89} for the PNLF, then $N_{0.5} =
N_{PN}/100$, where $N_{PN}$ is the total number of PNs associated with
the luminosity of the parent stellar population.  The
luminosity-specific PN number $\alpha = N_{PN} / L_{gal}$ is given by
\citet{Buzzoni06} for stellar populations with different ages and
metallicities, and calibrated using the PN population in the Local
Group, Leo group, and the Virgo and Fornax clusters. The maximal
theoretical value of $\alpha$ which gives the largest PN population
for a given luminosity is $ \alpha_{max} = 1 PN\times (1.85 \times
10^6 L_\odot)^{-1}$. This theoretical maximal value is independent of
metallicity, and also provides an upper limit to the observed $\alpha$
for the PN population in different galaxy types \citep{Buzzoni06}.

In the M87 halo, we have $N_{0.5} = 9\pm 3$, which gives a total
population of $N_{PN} = 900$ and a minimal bolometric luminosity of
the parent population of $ L_{gal} = N_{PN} / \alpha_{max} = 1.66
\times 10^9 L_\odot$ over a total surveyed area of 570 arcmin$^2$.  We
can thus derive a lower limit to the mean surface brightness in the V
band, $\mu_V = 28.6$ mag arcsec$^{-2}$, for a possible intermediate
age parent population. Comparing with $\mu_V = 28.3$ measured by
\citet{Mihos+05} at the position of the \citet{Williams+07} field, we
find that, to justify the number of bright PNs, at least $50$\% of the
stars in this field would need to come from a 1 Gyr
population.  However, the upper limit to the contribution of a younger
($< 10$ Gyr) component to the IRGB stars given by \citet{Williams+07}
is 30\% - 20\% . We must then conclude that there are not enough
intermediate age stars in the \citet{Williams+07} field to justify the
observed $N_{0.5}$.

What are the possible alternatives, if an intermediate age population
is not present? The \citet{Williams+07} results indicate that the
Virgo core population of stars is dominated by low metallicity stars
([M/H]$ \leq -1$) with ages $> 10$ Gyr; thus we may argue that the
halo of M87 may be also metal poor. In this case, the PNLF might have
a bright cutoff that is different from that of a metal rich
population. However \citet{Dopita92}, and PNLF observations in
metal-poor galaxies \citep{Ciardullo+02} show that metal poor systems
have values of $M^*$ that are fainter than those of their metal-rich
counterparts. Also the Oxygen abundance measurements by
\citet{Mendez05} of the brightest extragalactic PNs in NGC~4697
indicate near solar metallicities for these stars. Therefore PN
evolution from a metal poor population is unlikely to be a viable
explanation for the observed large core masses in the M87 halo PN.
The evolution of single stars from a 10 Gyr old population leads to
central star masses in the range $ 0.52 < $M$ < 0.55$ M$_\odot$
\citep{Buzzoni06}, which cannot supply the [OIII] 5007\AA\ flux
required at the PNLF bright cut-off of $M^*= -4.48$.  This led
\citet{Ciardullo+05} to propose an alternative form of evolution, i.e
close binaries and blue stragglers stars, as the likely progenitors of
[OIII] - bright PNs in a $10$ Gyr old stellar population. This
evolutionary channel seems better in agreement with the observations
for the brightest PNs in the M87 halo, than either the $\sim 1$
Gyr-old or the metal-poor progenitors.

\section{Conclusions}

We have measured the nebular [OIII] 5007\AA\ spectroscopic expansion
velocities for 11 PNs in the Virgo cluster core, at 15 Mpc distance,
with the FLAMES spectrograph on the ESO VLT. Based on the [OIII] line profile
width, the average spectroscopic expansion velocity for this
sample is $ \bar{v}_{HWHM} = 16.5$ kms$^{-1}$ (RMS = 2.6 kms$^{-1}$),
which is in agreement with the predictions of dynamically evolving
nebular models for high density nebulae close to their maximal
$m(5007)$ emission. Large central star masses M$_{CS} > 0.6$~M$_\odot$
are inferred from the bright measured [OIII] luminosities and the known
distance for the PNs bound to the M87 halo. From the large central
star masses and the measured $v_{HWHM}$, we derive short PN
visibility times and small nebular outer radii, $ \sim 0.07$ pc.  The
PNs in the M87 halo have large central star masses, are compact and
their nebula shells may be similar to those observed for the brightest
Galactic Bulge PNs.  Three mechanisms are reviewed as possible
explanation for the large core masses of the M87 halo PNs:
intermediate age population, metallicity effects, and blue stragglers,
with the latter being the most likely, given the old age and the low
metallicities of the IRGB stars in the Virgo core \citep{Williams+07}.

\acknowledgments

We thank the referee, Dr Detlef Sch\"onberner, for sharing his results
before publication and for his constructive comments. We thank Nando
Patat and Marina Rejkuba for support.

{\it Facilities:} \facility{ESO VLT}, \facility{HST (ACS)}.

\clearpage

\end{document}